# Cramer-Rao bound in detection of reversals

S.A.Ivanov[1], S.A.Merkuryev[1]

[1] St.Petersburg Branch of Pushkov Institute of Terrestrial Magnetism, Ionosphere and Radio Wave Propagation, 199164, St.Petersburg, Mendellevskaya, 1, Russia, e-mail: sergei.a.ivanov@mail.ru;

**Abstract.** We use the Cramer-Rao bound to show which short magnetic reversals can be recovered successfully from a magnetic anomaly profile. Using a simple variant of the Vine-Matthews model, we show, in particular, that errors in determining the duration of a short subchron are much smaller than errors in localizing the center of the subchron.

**Introduction.** Seafloor spreading magnetic anomalies give fundamental observations for detecting reversals of Earth's magnetic field and describing long-term plate motions in nearly all areas of the global oceans. Because identifications of magnetic reversals are usually accomplished by visual comparisons of real and modeled magnetic profiles, different interpretations of the same data occasionally occur depending on the experience of the interpreter. Differing or incorrect interpretations of one or several magnetic profiles may give rise to significantly different descriptions of plate motions for some areas of the mid-ocean ridges, and may also impact the reliability and accuracy of geomagnetic polarity time scales that rely in part on accurate identifications of magnetic reversals for age estimates of those reversals.

Here, we show that in the framework of the Vine-Matthews model for seafloor spreading [Vine-Matthews,1963] the Cramer-Rao bound can be applied to the related problems of detecting short-duration magnetic polarity intervals (i.e. subchrons) and evaluating the accuracy with which subchrons can be located and their durations determined. The full version will be published soon [Ivanov S.A., Merkuryev S.]

**Geophysical Model.** We consider the Vine-Matthews model [Vine-Matthews, 1963], in which magnetic anomalies are created by uniformly magnetized, long rectangular prisms of oceanic crust whose magnetism is given by the polarity state of the Earth's magnetic field at the time the seafloor is created. For a prism at the magnetic pole, parallel to the y-axis, with cross-section $\alpha < x < \beta$, $-b < z < -a$, and vertical magnetization $\mathbf{M}=M\mathbf{e}_z$ its associated magnetic anomaly field at height $z$ or depth $-z$ is

$$y(x,z) = -\frac{\mu_0 M_z}{2\pi}\left[\arctan\frac{x-\beta}{z+b} - \arctan\frac{x-\alpha}{z+b} - \arctan\frac{x-\beta}{z+a} + \arctan\frac{x-\alpha}{z+a}\right]. \quad (1)$$

**Cramer-Rao bound.** Suppose we have observations $\tilde{y}_i$ of the anomaly field $y_i$ at points $\tilde{x}_i$, with $z$ fixed and the field perturbed by noise $\tilde{y}_i = y_i + \varepsilon_i$. Here $\varepsilon_i$ are independent normally distributed random variables with zero mean and variance $\sigma^2$. Let $p(u_1,u_2,\ldots,u_n) = \frac{1}{\sqrt{(2\pi\sigma)^n}} e^{-\Sigma(u_i-y_i)^2/2\sigma^2}$ be the joint probability density and consider $l = \log(p)$ as a function of the observations $\tilde{y}_i = u_i$ and the geometrical parameters of the prisms. The quantity $I = E[(\partial_{X_1} l)^2]$ is called the Fisher information (E means the expectation). Suppose we have a method that recovers a parameter of the magnetized slab, say, the x-coordinate $\alpha$ of an edge of a prism, and also assume the method has no systematic error. Then the variance of the estimates $\text{Var}(\alpha) = \sigma^2(\alpha)$ satisfies (under additional conditions) the Cramer-Rao inequality $\sigma^2(\alpha) \geq 1/I$. If we estimate several parameters, the Fisher Information Matrix $I$ is written with the entries $E[\partial_\alpha l \partial_\beta l]$ and the Cramer-Rao inequality gives the estimate of the covariance matrix by $I^{-1}$, see [Borovkov,1984]. In particular, the low bounds of the variances of the estimates are the diagonal entries of $I^{-1}$.

**Example.** Consider a magnetized slab consisting of two prisms of positive (normal) magnetization separated by a narrow prism with negative (reverse) magnetization (Fig. 1). The x-coordinates of the edges are $\{X_1, X_2, X_3, X_4\} = \{12.644, 17.524, 18.134, 25.576\}$. The top of the basalt body is at depth 4 km and the bottom at 4.4 km. The magnetization is 10 A/m. To the field generated by these prisms we add normally distributed noise with zero mean and standard deviation of 10% of the magnitude of the field (equal to 63.7734 nT).

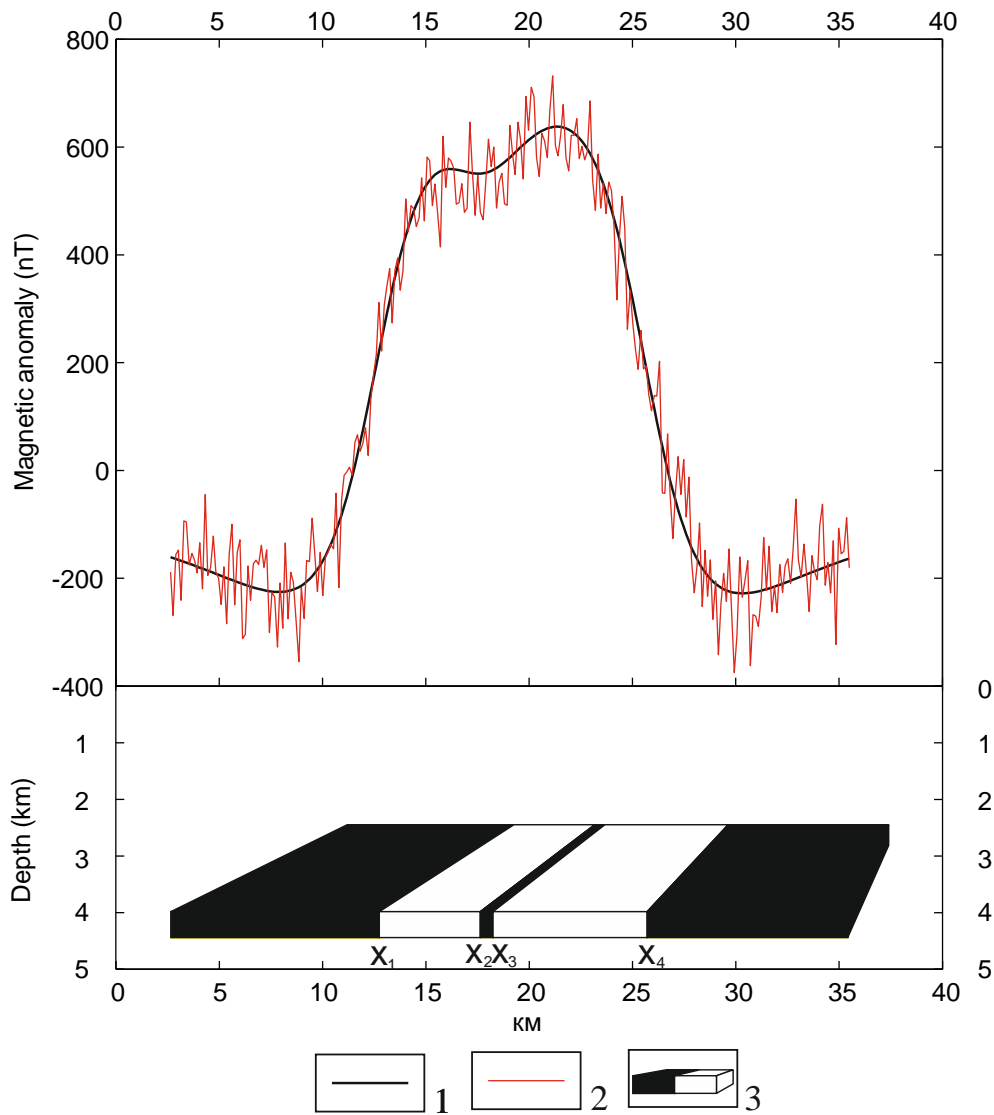

Fig. 1. The model (3) of the magnetized body, the field (1) and the field (2) with the noise on the surface.

**Estimates of the errors and scattering ellipse.** We wish to locate the narrow prism, i.e., to determine its coordinates $X_2$ and $X_3$. The specified width of this prism is $\delta = X_3 - X_2 = 0.61$ km and its center is $X_c = (X_3 + X_2)/2$. For the standard deviation, the Cramer-Rao inequality gives

$$\sigma_\delta \geq \sigma_\delta^{CR} = 0.1822 \text{ km}, \quad \sigma_c \geq \sigma_c^{CR} = 0.7283 \text{ km}.$$

To show the results for $X_2$ and $X_3$ we use the standard scattering ellipse.

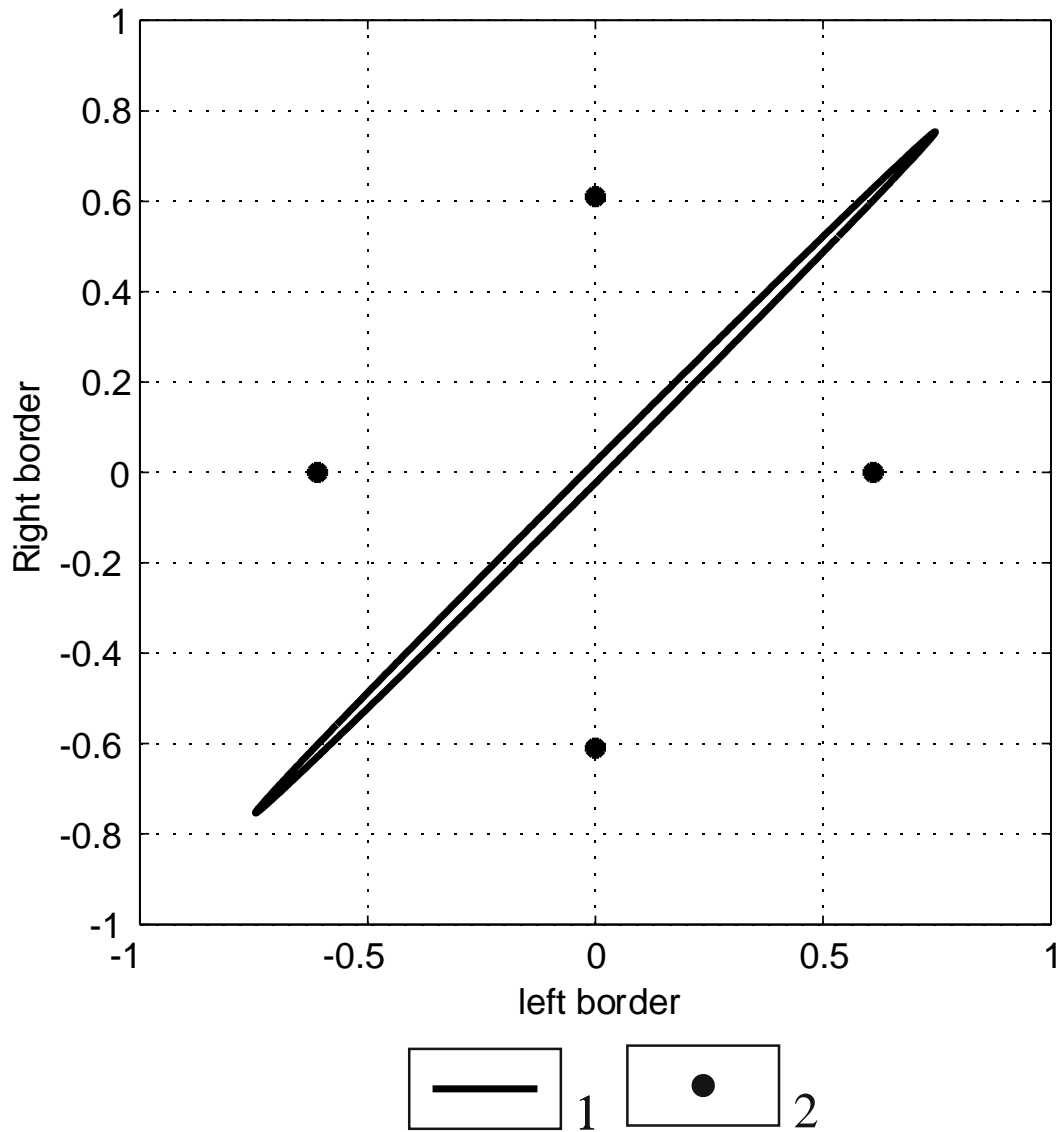

Fig.2. Scattering ellipse (1) for the ends $X_2-X_c$ and $X_3-X_c$ of the middle body. The points (2) with the coordinates $\pm \delta$ are shown where $\delta$ is the width of the body $\delta = X_3 - X_2$.

The probability that the estimate of the pair $X_2-X_c$ and $X_3-X_c$ lies within the ellipse is 40% or less. We see that the pair is strongly positive correlated and the width can be detected much more reliably than the ends.

*Remark 1.* The estimates of the deviations obtained by the least squares method are close to estimates from the Cramer–Rao bound. Because the specified noise is normally distributed, the least-squares and maximum-likelihood methods give similar results, with parameter error estimates that asymptotically approach the minimum errors.

**Near-bottom magnetometer observations.** As is well known, observing magnetic anomalies on the ocean surface is equivalent to applying a bandpass filter (an *earth filter)* that attenuates low and high wave-number components. In contrast, measurements closer to the magnet source (the ocean bottom) better resolve the details of magnetic anomalies. For example, Figure 3 shows the modeled magnetic fields at the ocean surface (dashed line) and at a depth of 3 km (solid line) generated by the three prisms.

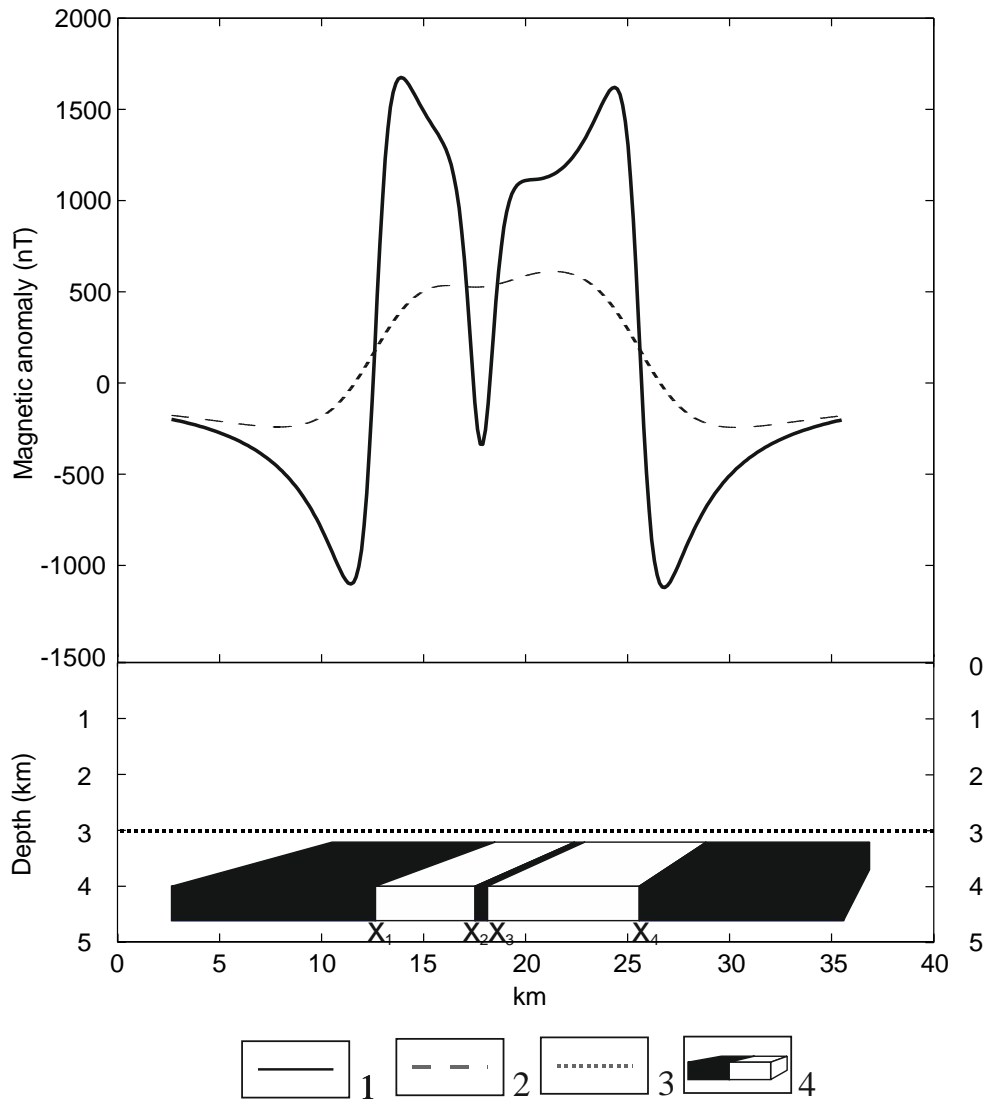

Fig.3. The model magnetic anomaly field at the ocean surface (dashed line (2) ) and the field (1) (solid line (1) at a depth of 3 km (the dotted line (3) in the lower panel). Prisms (4) in the lower panel show the magnetized body used to create the synthetic magnetic anomalies.

As expected, observations at a depth of 3 km, only 500 m above the top of the magnetized slab, resolve the reversal boundaries much better than the surface observations, with standard deviations from the Cramer–Rao approach of $\sigma_\delta \geq 0.0731$ km and $\sigma_c \geq 0.0788$ km.

**Spreading rate and detection of short magnetic reversals.** Figure 4 shows the magnetic anomalies calculated at the ocean surface for magnetic reversals A1 to A5 and four progressively faster half-spreading rates of 0.5 - 2.0 cm/yr. At the two slowest rates, the anomaly peaks converge and overlap, whereas most of the magnetic anomalies associated with the reversal are distinct at the fastest spreading rate.

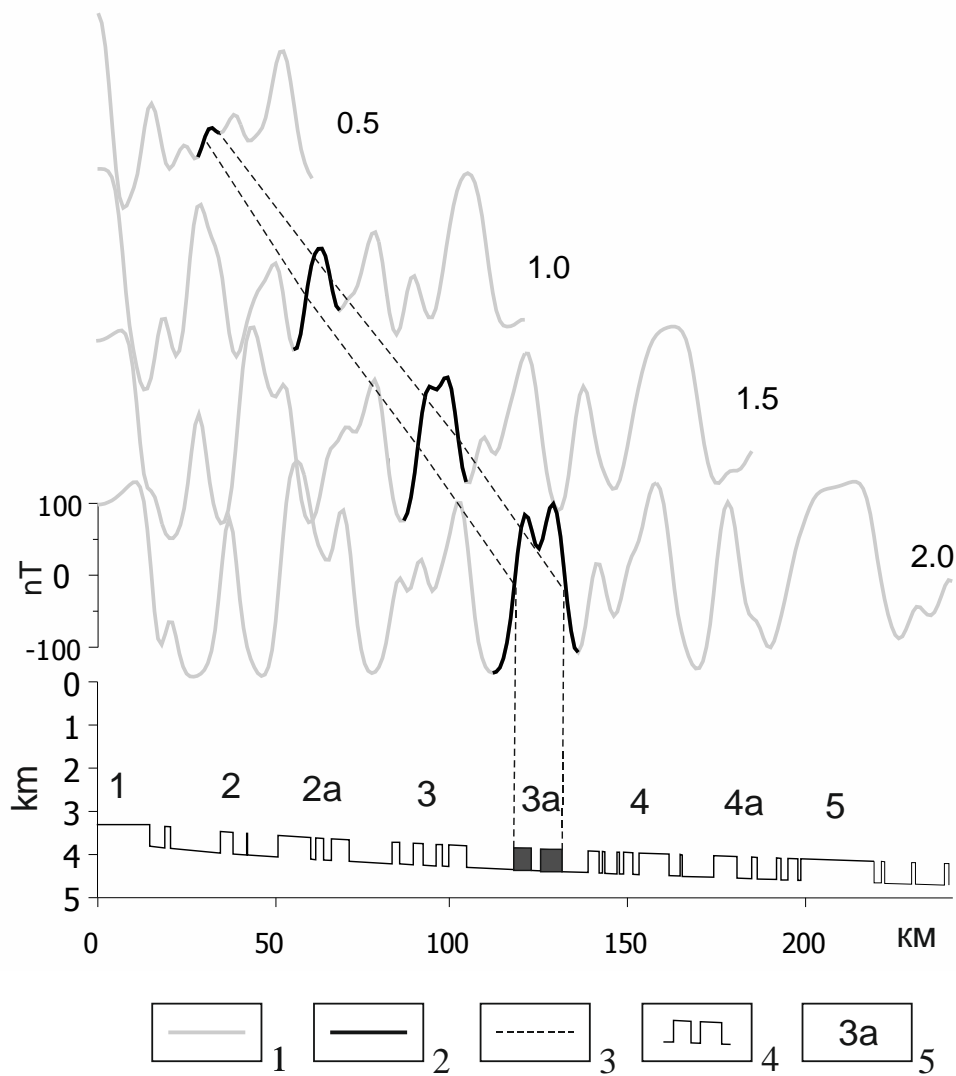

Fig.4. Field of anomalies for four progressively faster half- spreading rates.
(1) – model fields;
(2) – magnetic anomaly for Chron 3A, bold line;
(3) - magnetic anomaly for Chron 3A for each spreading rates, dashed line;
(4) – the Anomaly 1 to 5 reversal sequence used to create the synthetic anomalies
(5) – the names of the chrons.

For example, Anomaly 3A is observed along the Reykjanes Ridge in the north Atlantic, where the spreading rate is ~1.2 cm/yr. At this half-spreading rate, the width δ of the Anomaly 3A subchron should be 2.64 km. If we add normally-distributed noise to the above magnetic profile and plot the relative errors $\sigma_\delta^{CR}/\delta$ and $\sigma_c^{CR}/\delta$, where $\sigma_\delta^{CR}$ and $\sigma_c^{CR}$ are the Cramer-Rao (lower) bounds of the standard deviation of the width and of the centre of the anomalies correspondingly, Figure 5 shows that our ability to find the center location of this anomaly degrades far more rapidly at slower seafloor spreading rates than does our ability to resolve the width (duration) of the anomaly.

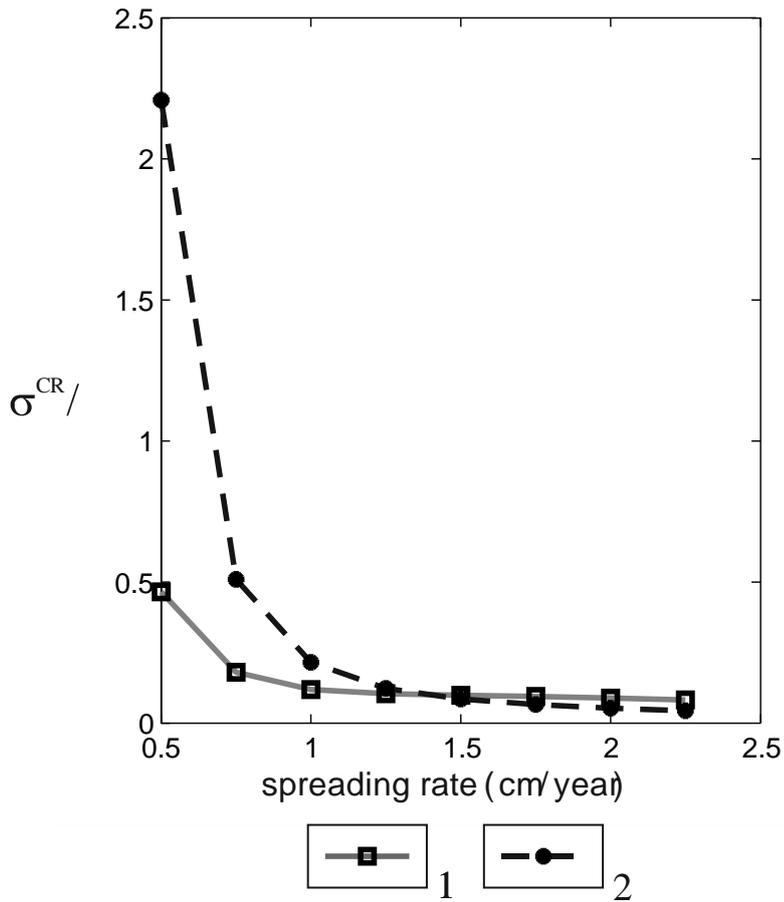

Fig. 5. (1) - relative errors $\sigma_\delta^{CR}/\delta$ of the width of the subchron, (2) - relative errors $\sigma_c^{CR}/\delta$ of the centre of the subchron.

*Remark 3.* In our noise model the Cramer-Rao bounds of the errors are related linearly with the standard deviation of the noise added to the anomaly field. If we take σ equal to 5% of the magnitude (instead of 10%), then the bounds $\sigma_\delta^{CR}$ and $\sigma_c^{CR}$ decrease by half.

**Future Problems.** 1. Consider seafloor spreading model(s) more complicated than a set of rectangular prisms.
2. Consider more realistic noise model for real observations, including quantization error.

**Conclusions.** 1. Cramer-Rao bounds can be applied to estimate error bounds for the location and duration of short magnetic polarity intervals.
2. The duration of a subchron can be determined more accurately than its centre.

**References.**